\documentclass[11pt]{article}
\usepackage{amssymb}
\usepackage{amsmath}
\usepackage{amsthm}

\usepackage{graphicx,color}

\newcommand{\ket}[1]{\left | #1 \right \rangle}
\newcommand{\bra}[1]{\left \langle #1 \right |}

\def\openone{\leavevmode\hbox{\small1\kern-3.8pt\normalsize1}}
\def\tr{{\rm Tr}\; }
\def\ch{{\cal H}}
\def\cm{{\cal M}}

\def\ce{{\cal E}}

\def\cH{{\cal H}}

\newcommand{\be}{\begin{equation}}
\newcommand{\ee}{\end{equation}}
\newcommand{\bea}{\begin{eqnarray}}
\newcommand{\eea}{\end{eqnarray}}
\newcommand{\beann}{\begin{eqnarray*}}
\newcommand{\eeann}{\end{eqnarray*}}

\def\RR{\mathbb{R}}

\newtheorem{theorem}{Theorem}

\newtheorem{proposition}{Proposition}
\newtheorem{corollary}{Corollary}
\newtheorem{conjecture}{Conjecture}
\theoremstyle{definition}

\newcommand{\proj}[1]{\ket{#1}\!\bra{#1}}

\newcommand{\vc}[1]{\underline{#1}}

\newcommand{\acc}{{I_{acc}}}

\def\be{\begin{equation}}
\def\ee{\end{equation}}
\def\bea{\begin{eqnarray}}
\def\eea{\end{eqnarray}}
\def\reff#1{(\ref{#1})}

\begin{document}
\title{\LARGE\bf
  Properties of subentropy}
\author{Nilanjana Datta$^1$, Tony Dorlas$^2$,
  Richard Jozsa$^3$
  and Fabio Benatti$^4$\\[3mm]
  \small\it $^1$Statistical Laboratory, Centre for Mathematical Sciences, University of Cambridge,\\ \small\it Wilberforce Road, Cambridge CB3 0WA, U.K.\\[1mm]
  \small\it $^2$School of Theoretical Physics, Dublin Institute for Advanced Studies,\\  \small\it 10 Burlington Road, Dublin 4, Ireland.\\[1mm]
  \small\it $^3$DAMTP, Centre for Mathematical Sciences, University of Cambridge,\\ \small\it Wilberforce Road, Cambridge CB3 0WA, U.K.\\[1mm]
  \small\it $^4$Dipartimento di Fisica, Universit\`a di Trieste, I-34151 Trieste, Italy \\  \small\it \&
Istituto Nazionale di Fisica Nucleare, Sezione di Trieste, I-34151 Trieste, Italy\\[1mm]
}

\date{}

\maketitle

\begin{abstract}
 Subentropy is an entropy-like quantity that arises in quantum information theory; for example, it provides a tight lower bound on the accessible information for pure state ensembles, dual to the von Neumann entropy upper bound in Holevo's theorem. Here we establish a series of properties of subentropy, paralleling  the well-developed analogous theory for von Neumann entropy. Further, we show that subentropy is a lower bound for min-entropy. We introduce a notion of conditional subentropy and show that it  can be used to provide an upper bound for the guessing probability of any classical-quantum state of two qubits; we conjecture that the bound applies also in higher dimensions. Finally we give an operational interpretation of subentropy within classical information theory.
\end{abstract}

\section{Introduction}\label{intro}
Subentropy is an intriguing entropy-like quantity that first appeared in \cite{ww1} and was named in \cite{jrw}. Let $\rho$ be any state (density matrix) of a quantum system
with an $n$-dimensional complex Hilbert space $\ch_n$ and let $\vc{\lambda}= (\lambda_1, \ldots ,\lambda_n )$ denote the eigenvalues of $\rho$, which we always list in non-increasing order $1\geq \lambda_1 \geq \ldots \geq \lambda_n \geq 0$. Then the subentropy $Q(\rho )$ of $\rho$ is the function of its eigenvalues given by\footnote{In this paper, all entropic quantities are defined in terms of natural logarithms.}
\begin{equation} \label{qdef}
Q(\rho)=F(\vc{\lambda}) := - \large\sum_{i=1}^n  \frac{\lambda_i^n}{ \prod_{ j \ne i}(\lambda_i - \lambda_j)}\, \ln \lambda_i .
\end{equation}
If eigenvalues coincide (or are zero) we define $Q(\rho)$ to be the corresponding limit of the above expression, which is always well-defined and finite. For example if $\lambda_1=\lambda_2$ then the $i=1,2$ terms of eq.(\ref{qdef}) are singular but in the limit as $\lambda_2 \rightarrow \lambda_1$, these terms taken together simply construct the derivative of $f(x)=x^n\ln x$ at $x=\lambda_1$ via its differential quotient.

The expression in eq.(\ref{qdef}) arose in \cite{ww1} in the context of the following information theoretic issue. Let $\cm$ denote a complete von Neumann measurement associated to an orthonormal basis $\{ \ket{e_i} \}$ in $\ch_n$. If $\cm$ is applied to $\rho$ we get the post-measurement state $\cm (\rho)= \sum_i p_i \proj{e_i}$ with outcome probabilities given by $p_i=\bra{e_i}\rho\ket{e_i}$. Let $H(\cm (\rho ))= -\sum_i p_i \ln p_i$ denote the Shannon entropy of the output distribution. Now consider the average of this entropy over all choices of von Neumann measurements. More precisely, orthogonal bases in $\ch_n$ are related by unitary transformations and we average over all choices of basis $\{ \ket{e_i} \}$ with respect to the Haar measure on the unitary group. Denoting the Haar average by $\langle \,\cdots\, \rangle_\cm$ it was shown in \cite{ww1} (with further calculational details in \cite{jrw}) that
\begin{equation} \label{aveh} Q(\rho)=\left\langle H(\cm (\rho )) \right\rangle_\cm - C_n \hspace{5mm}
 \mbox{where} \hspace{3mm} C_n := \frac{1}{2}+ \frac{1}{3} +\ldots +\frac{1}{n}.
 \end{equation}
To develop a further useful expression for $Q(\rho)$  let $\{ \ket{\lambda_k} \}$ be the orthonormal basis of eigenvectors of $\rho$ and for the varying bases $\{ \ket{e_i} \}$ consider each basis vector expanded in the eigenbasis:
\[ \ket{e_i}=\sum_k a_k\ket{\lambda_k}, \hspace{3mm} \mbox{and write}\hspace{3mm} x_k=|a_k|^2,  \]
so that \[ p_i=\bra{e_i}\rho\ket{e_i}=\sum_k\lambda_k x_k=  \vc{\lambda}\cdot\vc{x}. \]
Now as the basis varies under the Haar distribution, each basis vector $\ket{e_i}$ becomes Haar-uniformly distributed over $\ch_n$, and according to a result of Sykora \cite{sykora} the corresponding vector $\vc{x}$ of squared coefficients is then distributed {\em uniformly} over the $(n-1)$-dimensional probability simplex \[ \Delta_n = \{ \vc{x}=(x_1, \ldots , x_n): x_i \geq 0 \mbox{ and } \sum_i x_i=1 \},\]
Using $(x_1, \ldots , x_{n-1})$ as co-ordinates in $\Delta_n$, this uniform measure (normalised to have total volume unity) is given by
\begin{equation}\label{unifmeas}
d {\bf x} = (n-1)!\,\,  dx_1 \ldots dx_{n-1}.
\end{equation}
Returning to our average of the Shannon entropy in eq.(\ref{aveh}), we see that each term $(-p_i\ln p_i)$ contributes the same average.
Introducing
\[ \eta (y) := -y\ln y, \]
we then obtain a third expression for the subentropy from eq.(\ref{aveh}):
\begin{equation} \label{etaave}  Q(\rho ) = n \int_{\Delta_n}\! \eta ( \vc{\lambda}\cdot \vc{x})\,\, d{\bf x} - C_n. \end{equation}
In \cite{jrw} an operational meaning of subentropy was established in terms of the notion of accessible information.
Let $\ce= \{q_i, \ket{\psi_i}\}$ be an ensemble of pure states with density matrix $\rho = \sum_i q_i \proj{\psi_i} $.
We refer to any such ensemble
as a $\rho$-ensemble. The accessible information $I_{acc}(\ce)$ of $\ce$ is the maximum amount of classical information about the value of $i$ that can be obtained by
any quantum measurement on the pure states $|\psi_i\rangle$.
According to Holevo's theorem \cite{holevo},  the von Neumann entropy $S(\rho ) = -\tr( \rho \ln\rho) $ is a tight upper bound on $\acc (\ce )$ for all $\rho$-ensembles, being attained for the eigenstate ensemble $\{\lambda_i, \ket{\lambda_i}\}$.
In \cite{jrw} it was shown that, dually, the subentropy $Q(\rho )$ is a tight {\em lower} bound on $\acc (\ce )$ as $\ce$ ranges over all $\rho$-ensembles, being attained for the so-called Scrooge ensemble \cite{jrw}. In particular, this implies that $Q(\rho ) \leq S(\rho )$ for all $\rho$.


Some further basic properties of the subentropy were established in \cite{jrw}. If we write the subentropy of a state $\rho$
as $Q(\rho)=-\sum_i c_i \ln \lambda_i$ (with the eigenvalues $\lambda_i$ arranged in non-increasing order) then curiously, as for the Shannon entropy,
the coefficients $c_i$ satisfy $\sum_i c_i=\sum_i \lambda_i$, but they alternate in sign (for the generic case of distinct eigenvalues) and they can be of unbounded magnitude. From eq.(\ref{etaave}) (remembering that $\eta$ is a strictly concave function) we see that $Q(\rho)$ (or more precisely the function $F(\underline{\lambda})$ in eq.(\ref{qdef})) is a strictly concave function of the $\lambda_i$'s. Since it is also symmetric, it must attain its maximum in dimension $n$ when all eigenvalues are equal i.e. $\lambda_i=1/n$, and the maximum is given by
\[ Q\left(\frac{I}{n} \right)= \ln n - C_n. \]
This is monotonically increasing with $n$ and bounded above by
\[ \lim_{n\rightarrow \infty}\,\,  \ln n - C_n= 1-\gamma \approx 0.42278 \]
where $\gamma$ is Euler's constant. Thus for any state $\rho$ on any $n$-dimensional Hilbert space, the subentropy $Q(\rho )$ is upper-bounded by 0.42278, whereas $S(\rho )$ may be as large as $\ln n$.

For pure states the subentropy is zero (e.g. as $Q(\rho )\leq S(\rho )$) and setting $\vc{\lambda} = (1,0, \ldots , 0)$ in eq.(\ref{etaave}) we obtain the useful integral
\begin{equation} \label{intxlnx}
-\, n\int_{\Delta_n}\! x_1 \ln x_1\,\, d{\bf x} = C_n,
\end{equation}
(which may also be evaluated directly by elementary means).

Even though  subentropy was introduced more than a decade ago, it has remained largely unexplored. It has appeared in \cite{jacobs} to provide bounds on information gain from efficient quantum measurements, and in \cite{jm04} in the study of quantum  information compression. In \cite{nichols} a series of quantities interpolating between entropy and subentropy was studied and in \cite{mintert} a notion of Renyi subentropy was introduced.

In this paper we derive a series of properties of subentropy, which are analogous to those of von Neumann entropy. We also prove that it provides a lower bound to the more recently defined {\em{min-entropy}} \cite{renner}, which plays a pivotal role in one-shot information theory.
Numerical investigations suggest that subadditivity, which is a fundamental property of von Neumann entropy, is also valid for subentropy. We provide
an analytical proof of this for product states.

We also introduce a notion of conditional subentropy (analogous to conditional von Neumann entropy) and conjecture that a lower bound on an interesting operational quantity called the {\em{guessing probability}} can be expressed in terms of it. This conjecture is supported by an analytical proof for $n=2$, and numerical evidence for $n=3$.

The expression for subentropy given by the function $F$ in eq.\reff{qdef} may be applied to any probability distribution, and it is interesting to ask
whether it can be given an operational meaning within {\em{classical}} information theory. We answer this question in the affirmative by
proving that the subentropy of a discrete random variable $X$ is equal to the mutual information between the input and output of a specific classical channel, when
$X$ is considered as the random variable characterizing the input. Consequently the subentropy also provides a lower bound on the capacity of such a channel.

\section{Properties of subentropy}\label{sect2}
\subsection{Subentropy as an averaged quantum relative entropy}\label{relent}
The quantum relative entropy of a state $\rho$ and a positive semi-definite operator $\sigma$ is defined as
\begin{equation}
 D(\rho || \sigma ) = \left\{ \begin{array}{cc} \tr (\rho \ln \rho - \rho \ln \sigma )  & \quad {\hbox{if }} {\rm{supp}}\,  \rho \subseteq {\rm{supp}}\, \sigma\nonumber\\
 \infty &\quad {\hbox{otherwise}}, \end{array} \right.
\end{equation}
where ${\rm{supp}} \, \rho$ denotes the support of $\rho$.
Let $\rho = \sum_j p_j P_j$ be any convex decomposition of $\rho$ into pure states $P_j = \proj{\phi_j}$. Let $\cm (\rho )$ and $\cm ( P_j)$ denote the post-measurement state after a complete von Neumann measurement $\cm$ on the states $\rho$ and $P_j$, respectively.

\begin{proposition} \label{proprelent} The subentropy of $\rho$ is given by
\be\label{rel1} Q(\rho) = \sum_j p_j \left\langle D(\cm (P_j )|| \cm (\rho )) \right\rangle_\cm,
\ee
where $\langle \cdots \rangle_\cm$ denotes the Haar measure average over choice of measurement basis.
\end{proposition}
\noindent {\bf Proof}\, We have
\[  \sum_j p_j \left\langle D(\cm (P_j )|| \cm (\rho )) \right\rangle_\cm =
\sum_j p_j \tr \cm (P_j)\ln \cm (P_j) - \sum_j p_j \tr \cm (P_j)\ln \cm (\rho ). \]
Since $\rho = \sum_j p_j Pj$ we have $\sum_j p_j \cm (P_j)= \cm (\rho )$, so the second term above is $H(\cm (\rho ))$ with average $\langle H(\cm (\rho )\rangle_\cm $. In the first term each $P_j$ is a pure state and the Haar average over $\cm$ is independent of the choice of state, giving the same result for each $j$. Inserting the pure state diag$(1,0, \ldots , 0)$ for $P_j$ and mapping the Haar average to an integral over the probability simplex (via Sykora's theorem \cite{sykora}) we can directly use eq.(\ref{intxlnx}) to get
\[  \sum_j p_j \left\langle D(\cm (P_j )|| \cm (\rho )) \right\rangle_\cm = -C_n+ \langle H(\cm (\rho )\rangle_\cm, \]
which equals $Q(\rho )$ by eq.(\ref{aveh}).\, $\Box$

As an application of this result we get a simple proof that $Q(\rho )\leq S(\rho )$ (which is not easy to see directly from the other formulae, eqs.\reff{qdef}, \reff{aveh} and \reff{etaave}). Indeed the relative entropy is well known to be non-increasing under quantum operations \cite{nc} such as measurements $\cm$. Thus $Q(\rho ) = \sum_j p_j \langle  D(\cm (P_j )|| \cm (\rho )) \rangle_\cm \leq \sum_j p_j \langle  D(P_j || \rho ) \rangle_\cm = S(\rho )$ (where the last equality follows since $D(P_j || \rho )$ is independent of $\cm$ and $\rho = \sum_j p_j P_j$).

\subsection{Concavity and Schur concavity of subentropy}\label{concave}
We noted above that the subentropy formula eq.(\ref{qdef}) is concave as a function of the $\lambda_i$'s. We show here that even more, $Q(\rho )$ is concave and also Schur concave, as a function of the quantum state $\rho$ (which are also properties of von Neumann entropy).
\begin{theorem}\label{op_concave}
The subentropy $Q(\rho)$ is concave as a function of the density
operator $\rho$, i.e.
 if
$\rho_1,\dots,\rho_n$ are density matrices and $\{q_i\}_{i=1}^n$
is a probability distribution, then
\be\label{concav} Q\left(\sum_{i=1}^n q_i \rho_i \right) \ge
\sum_{i=1}^n q_i Q(\rho_i). \ee
\end{theorem}
\noindent {\bf Proof}\,
If $\cm$ is any complete von Neumann measurement then, by concavity of the entropy,
\[
H\bigl({\cm}\bigl( \sum_i q_i\rho_i\bigr)\bigr) = H\bigl( \sum_i q_i \cm(\rho_i)\bigr)
\ge \sum_i q_i H\bigl(\cm(\rho_i)\bigr).\]
Using eq.(\ref{aveh}) and the above inequality, we obtain
\bea
Q\bigl(\sum_i q_i \rho_i\bigr) &=& \langle H\bigl({\cm}(\sum_i q_i \rho_i)\bigr)\rangle_\cm - C_n \nonumber\\
&\ge  & \langle\sum_i q_i H\bigl(\cm(\rho_i)\bigr)\rangle_{\cm} - C_n\nonumber\\
&=& \sum_i q_i \bigl(\langle H\bigl(\cm(\rho_i)\bigr)\rangle_{\cm} - C_n\bigr)\nonumber\\
&=& \sum_i q_i Q(\rho_i).\hspace{2cm} \Box\nonumber
\eea
We remark that the result of Theorem \ref{op_concave} is actually implicit in \S V of \cite{jrw} where it is shown that for any ensemble $\ce=\{p_i, \rho_i\}$ of {\em mixed} states, with $\sum_ip_i\rho_i=\rho$, the quantity  $Q(\rho)-\sum_i p_i Q(\rho_i)$ is the average classical information obtainable about $i$ from complete von Neumann measurements
$\cm$ (where we average over all choices of $\cm$).

\begin{corollary} \label{mixit} Subentropy is non-decreasing under mixing-enhancing maps i.e.~if $\{U_i\}_i$ is a set of unitary operators and $\{ q_i\}_i$ is a probability distribution, then
\[ Q\left(\sum_i q_i U_i\rho U^\dagger_i \right)\geq Q(\rho ). \]
More generally the $U_i$'s may range over a continuous distribution here.
\end{corollary}
\noindent {\bf Proof}\, Immediate from the concavity of $Q(\rho )$ and its invariance under unitary transformations: $Q(U\rho U^\dagger )=Q(\rho )$).\, $\Box$

Let $\rho$ and $\sigma$ be quantum states with eigenvalues $\vc{\lambda} = (\lambda_1, \ldots , \lambda_n)$  and $\vc{\mu} = (\mu_1, \ldots , \mu_n)$ respectively, listed in non-increasing order. We say that $\rho$ is majorised by $\sigma$ and write $\rho \prec \sigma$, if $\vc{\lambda}$ is majorised by $\vc{\mu}$ (in the usual sense of majorisation of vectors) i.e. if \[ \sum_{i=1}^k \lambda_i \leq \sum_{i=1}^k \mu_i \hspace{5mm} \mbox{ for all}\hspace{3mm} 1\leq k\leq n. \]
A real-valued function $F(\rho )$ is called Schur concave in $\rho$ if $\rho \prec \sigma \implies F(\rho ) \geq F(\sigma )$.
\begin{corollary}\label{schur} The subentropy $Q(\rho )$ is Schur concave in $\rho$. \end{corollary}
\noindent {\bf Proof}\, According to a theorem of Uhlmann (cf. \cite{wehrl} \S II\,C) $\rho \prec \sigma$ if and only if  $\rho = \Lambda(\sigma )$ for some mixing-enhancing map $\Lambda$. Then $Q(\rho )\geq Q(\sigma )$  follows immediately from Corollary \ref{mixit}.\, $\Box$

Schur concavity of subentropy was also noted in \cite{mintert}. By considering the special case of diagonal states, we infer from Corollary \ref{schur} that the function $F(\vc{\lambda})$ on $\RR^n$ given in eq.(\ref{qdef}) is also Schur concave as a function of $\vc{\lambda}$. Alternatively, the Schur concavity of $F$ follows from the fact that any function on $\RR^n$ that is symmetric and concave, is also Schur concave.

\subsection{Continuity of subentropy}
We prove that the subentropy $Q(\rho)$ satisfies a
continuity bound formally identical to  Fannes' inequality \cite{fannes}
for the von Neumann entropy $S(\rho)$.
\begin{proposition}\label{fanineq}
\label{continuity} Suppose states $\rho$ and $ \sigma$ on an $n-$dimensional Hilbert space have trace distance $D(\rho,
\sigma):= \frac{1}{2} ||\rho -
\sigma||_1=T$ with $ T\le 1/2e.$
Then
\be |Q(\rho) - Q(\sigma)| \le
2T \ln n + \eta\left(2T\right), \ee where
$\eta(y) = - y \ln y.$
\end{proposition}
\noindent {\bf Proof}\,  This follows by applying Fannes' inequality for the von
Neumann entropy to the formula eq.(\ref{aveh}). First
note that, since the trace norm is monotone under measurement
operations, we have
$$ || {\cm}(\rho) - {\cm}(\sigma)||_1 \leq || \rho - \sigma
||_1. $$ Then recalling that $\eta(y)$ is increasing
for $y \leq 1/e$, we get
\begin{eqnarray*} |Q(\rho) - Q(\sigma)| &=& |
\langle S({\cm}(\rho)) - S({\cm}(\sigma)) \rangle_{\cm}|
\\ &\leq & \langle | S({\cm}(\rho)) - S({\cm}(\sigma))| \rangle_{\cm}
\\ &\leq & \langle || {\cm}(\rho) - {\cm}(\sigma)||_1 \ln n
+ \eta \left( || {\cm}(\rho) - {\cm}(\sigma)||_1 \right)
\rangle_{\cm} \\ &\leq & || \rho - \sigma||_1 \ln n + \eta
\left( || \rho - \sigma||_1 \right). \hspace{2cm} \Box \end{eqnarray*}

In a similar way we can also derive an analogue of Audenaert's inequality \cite{koenraad} 
viz. with $\rho$, $\sigma$ and trace distance $T$ as in Proposition \ref{fanineq} above, we have
 \be |Q(\rho) - Q(\sigma)| \le T \ln (n-1) + h(T)
\ee
where $h(T)$ is the binary entropy function $h(T)= - T \ln T- (1-T) \ln (1-T)$.

\subsubsection{Is subentropy subadditive?}
A fundamental property of von Neumann entropy is subadditivity viz. $S(\rho_{AB})\leq S(\rho_A)+S(\rho_B)$ where $\rho_{AB}$ is a bipartite state with reduced states $\rho_A$ and $ \rho_B$. For product states $\rho_{AB}=\rho_A\otimes \rho_B$ it is easy to see directly that equality holds. For subentropy numerical investigations lead us to make the following conjecture:

\begin{conjecture} Subentropy is subadditive: $Q(\rho_{AB})\leq Q(\rho_A)+Q(\rho_B)$. \end{conjecture}

We prove the conjecture for the case of product states, where unlike von Neumann entropy, equality does not hold in general. Indeed the universal upper bound $1-\gamma \approx 0.42278$ on subentropy already shows that equality cannot hold for products of mixed states.
\begin{proposition}
$Q(\rho_{A}\otimes \rho_B)\leq Q(\rho_A)+Q(\rho_B)$.
\end{proposition}
\noindent {\bf Proof}\,  Let $\ce_A$, $\ce_B$ and $\ce_{AB}$ be the Scrooge ensembles \cite{hjw} of $\rho_A$, $\rho_B$ and $\rho_A\otimes \rho_B$. Thus the subentropies are given by the corresponding accessible informations \[ Q(\rho_A)= {\acc}(\ce_A),\hspace{5mm}Q(\rho_B)={\acc}(\ce_B),\hspace{5mm} Q(\rho_A\otimes \rho_B)={\acc}(\ce_{AB}). \]
Now $\ce_A\otimes \ce_B$ (the ensemble of all products of states from $\ce_A$ with those from $\ce_B$, taken with the corresponding product probabilities) is a $(\rho_A\otimes \rho_B)$-ensemble. Hence ${\acc}(\ce_A\otimes \ce_B) \geq {\acc}(\ce_{AB})$ (as the Scrooge ensemble has the least accessible information amongst all $\rho_A\otimes \rho_B$-ensembles). On the other hand it is known \cite{accinfo} that the accessible information is additive for tensor products of ensembles, so
\[ Q(\rho_A)+Q(\rho_B)= {\acc}(\ce_A)+ {\acc}(\ce_B)={\acc}(\ce_A\otimes \ce_B) \geq {\acc}(\ce_{AB})=Q(\rho_A\otimes \rho_B). \hspace{5mm}\Box \]

\subsection{Subentropy and min-entropy}

For any state $\rho$ the min-entropy
$H_{\min}(\rho)$ \cite{renner} is defined as\footnote{The min-entropy is usually defined in terms of logarithm to the base $2$ but here we use natural
logarithms.}: \be\label{min}
H_{\min}(\rho) := - \ln \lambda_{\max}(\rho), \ee where $
\lambda_{\max}(\rho)$ denotes the maximum eigenvalue of
$\rho$.  It is clear that $H_{\min}(\rho) \le S(\rho)$, where
$S(\rho)$ is the von Neumann entropy.
\begin{proposition}
For any state $\rho$ the subentropy is a
lower bound for the min-entropy:
\be\label{eq_thm1}
Q(\rho) \le H_{\min}(\rho).
\ee
\end{proposition}
\noindent {\bf Proof}\,  Let $\vc{\lambda}=(\lambda_1, \ldots , \lambda_n)$ be the eigenvalues of $\rho$ with $\lambda_1$ being the largest. Since $\sum_i \lambda_i=1$ we have $\lambda_1=1-\sum_{i=2}^n \lambda_i$, and $H_{\min }(\rho ) = -\ln (1-\sum_{i=2}^n \lambda_i)$ depends on $\lambda_2, \ldots , \lambda_n$ only through their sum. On the other hand from eq.(\ref{etaave}) we have
\[ Q(\rho) =-n \int_{\Delta_n}\! (\vc{\lambda}\cdot \vc{x})\ln ( \vc{\lambda}\cdot \vc{x}) \, d{\bf x} -C_n .\]
Setting $\lambda_1=1-\sum_{i=2}^n \lambda_i$ we see that $Q(\rho)$ is symmetric in $\lambda_2, \ldots , \lambda_n$ and then by concavity, for any given $\lambda_1$,  the maximum value $Q_{\max } $ occurs when $\lambda_2= \ldots = \lambda_n$. Writing $\lambda$ for this common value we have
\be\label{qmax}
Q_{\max }(\lambda) =-n \int_{\Delta_n}\! (\vc{\lambda}\cdot \vc{x})\ln ( \vc{\lambda}\cdot \vc{x}) \, d{\bf x} -C_n\hspace{5mm} \mbox{with $\vc{\lambda}=(1\! -\! (n-1)\lambda,\,  \lambda, \ldots , \lambda)$.} \ee
Also $H_{\min } (\rho)=-\ln (1\! -\! (n-1)\lambda )$, and so to show that $Q(\rho)\leq H_{\min }(\rho)$ it suffices to show that
$Q_{\max } (\lambda)\leq -\ln (1\! -\! (n-1)\lambda )$. Now $f(\lambda) =-\ln (1\! -\! (n-1)\lambda )$ satisfies
\[ \mbox{$f(0)=0$, \hspace{3mm} $f'(0)=(n-1)$\hspace{3mm} and $f$ is {\em convex} in $\lambda$}. \]
On the other hand $Q_{\max }$ satisfies
\[ \mbox{$Q_{\max }(0)=0$, \hspace{3mm}$Q_{\max}'(0)= (n-1)$ (as shown below), \hspace{3mm} and $Q_{\max }$ is {\em concave} in $\lambda$.} \]
Thus $f$ and $Q_{\max }$ have the same value and slope at $\lambda =0$. Also $f$ is convex while $Q_{\max}$ is concave, which implies that $Q_{\max}(\lambda) \leq f(\lambda)$, so $Q(\rho) \leq H_{\min }(\rho)$ as required.

To see that $Q_{\max}'(0)= (n-1)$ consider eq.(\ref{qmax}) with $\vc{\lambda}\cdot \vc{x}= (1\! -\! (n-1)\lambda)x_1+\lambda (x_2+\ldots +x_n)$. Then differentiating under the integral (and using $x_1+\ldots +x_n=1$) we get
\[ Q_{\max}'(\lambda)= -n\int_{\Delta_n}\! (1+\ln (\vc{\lambda}\cdot\vc{x}))\, (1-nx_1)\, d{\bf x}. \]
At $\lambda=0$ we have  $ \vc{\lambda}\cdot\vc{x}=x_1$, so
\[ Q_{\max}'(0)= -n\int_{\Delta_n}\! (1+\ln x_1)\, (1-nx_1)\, d{\bf x}. \]
This integral can be done by elementary means (e.g. using eq.(\ref{intxlnx}) and $\int_{\Delta_n}\! \ln x_1 \, d{\bf x}= -(1+ \frac{1}{2}+ \ldots + \frac{1}{n-1})$) to get $Q_{\max}'(0)= (n-1)$, completing the proof.\, $\Box$

\subsection{Conditional subentropy for classical-quantum states}
For a bipartite state $\rho_{AB}$ with subsystems $A$ and $B$ we define the conditional subentropy to be
\be\label{cond2}
Q(A|B)_\rho:= Q(\rho_{AB})-Q(\rho_B),
\ee
where $\rho_B$ is the reduced state of $B$ in $\rho_{AB}$.

A classical-quantum (c-q) state $\rho_{XB}$ is defined to be any state on a Hilbert space $\cH_X \otimes \cH_B$ of the form
\be\label{cq} \rho_{XB}= \sum_{x=1}^n p_x \proj{x}\otimes \rho^x_B \ee
where $\{ p_x \}$ is a probability distribution, $\{ \ket{x} \}$ is a set of orthonormal states in $\cH_X$, (with $x$ denoting the values
taken by a random variable $X$ with probability mass function $p_x$) and $\rho^x_B$ are any associated states of $B$.  Although $Q(A|B)_\rho$ may in general be negative (e.g. when $\rho_{AB}$ is an entangled pure state) we have:
\begin{proposition}\label{posrel} If $\rho_{XB}$ is a c-q state then the conditional subentropy $Q(X|B)_\rho$ is non-negative.\end{proposition}
\noindent {\bf Proof}\, For the c-q state eq.(\ref{cq}) let $\rho^x_B$ have eigenvalues $\mu_{xi}$ and eigenstates $\ket{\mu_{xi}}$ with $x=1, \ldots ,n$ and $i=1, \ldots ,m={\rm dim}\,\cH_B$. Note that $\rho_{XB}$ is a block-diagonal matrix with $p_x\rho^x_B$'s on the diagonal, so its eigenvalues are $\{ p_x\mu_{xi} \}$ for all $x,i$ (and $x$ not summed). Next note that $\rho_B=\sum_x p_x \rho^x_B$ is a mixture of the pure states $\ket{\mu_{xi}}$ with probabilities $p_x\mu_{xi}$ respectively. In other words,
$\{p_x\mu_{xi}, \ket{\mu_{xi}}\}$ is a $\rho_B$-ensemble. Let $\{ \lambda_k \}$ be the eigenvalues of $\rho_B$.

Let $\vc{q}$ be the vector with $mn$ entries containing the $p_x\mu_{xi}$ in non-increasing order. Then $Q(\rho_{XB})=F(\vc{q})$ with $F$ as in eq.(\ref{qdef}).
Let $\vc{\lambda}$ be the vector with $mn$ entries containing the $\lambda_k$'s in non-increasing order,  and then padded with extra zeroes. Then $Q(\rho_B)=F(\vc{\lambda})$.

Now we know \cite{hjw} that $\vc{q}=A\vc{\lambda}$ where $A$ is a doubly stochastic matrix. (Indeed this holds in general for eigenvalues of any mixed state $\sigma$ and probabilities in any pure state $\sigma$-ensemble cf eq.(15) in \cite{hjw}). Thus  \cite{bhatia} $\vc{q}$ is majorised by $\vc{\lambda}$. Then the Schur-concavity of the subentropy (Corollary \ref{schur}) gives $F(\vc{q})\geq F(\vc{\lambda})$, so $Q(X|B)_\rho\geq 0$.\, $\Box$

\subsubsection{Guessing probabilities and min-conditional entropy}\label{mincond}
For any c-q state $\rho_{XB}$, the guessing probability $p_{\rm guess} (X|B)$ is defined as the maximum probability that a (generalised POVM) measurement on the system $B$ yields the correct value of the random variable $X$. It is given by
\[ p_{\rm guess}(X|B)= \max_{\{E_x\}: {\rm{POVM}}}\sum_x p_x \tr\! (E_x\, \rho^x_B),\]
where the maximization is over all possible POVMs: $\{E_x:\,E_x \ge 0 \, \forall\, x$; $\sum_x E_x = I\}$.
K\"onig, Renner and Schaffner \cite{KRS} proved that the guessing probability is given in terms of an entropic quantity of the c-q state $\rho_{XB}$, known as the min-conditional entropy \cite{renner} which is defined as follows:
\be\label{min-cond} H_{\min}(X|B) := \max_{\sigma_B
} \left\{- D_{\max}(\rho_{XB} || I_X \otimes
\sigma_B)  \right\}, \ee where for any state $\rho$ and a positive
operator $\sigma$, $D_{\max}(\rho||\sigma)$ denotes the
max-relative entropy which is defined as follows \cite{min-max}\footnote{In \cite{min-max} this quantity was defined in terms of logarithm to the base $2$ but here we choose natural logarithms.}: \be\label{infgamma}
D_{\max}(\rho||\sigma):= \inf \{ \gamma : \rho \le e^\gamma
\sigma\}. \ee
They proved that \cite{KRS}:
\be\label{hminkrs} H_{\min} (X|B) = - \ln p_{\rm guess} (X|B). \ee
Analogous to our previous result that $Q(\rho) \leq H_{\min} (\rho)$ for non-conditional quantities, we conjecture that $Q(X|B)$ is a lower bound for $H_{\min}(X|B)$. This would be a nontrivial lower bound as the conditional subentropy was shown above to be non-negative for all c-q states.
\begin{conjecture} For a c-q state $\rho_{XB}$, the conditional subentropy $Q(X|B)$ satisfies the following bound:
\[ {Q(X|B)} \leq H_{\min} (X|B) = -\ln p_{\rm guess}(X|B).\hspace{1cm} \Box \]
\end{conjecture}
The quantity $H_{\min} (X|B)$ (or $p_{\rm guess}(X|B)$) is not readily computable from its definition whereas $Q(X|B)$ is directly computable. A significant consequence of our conjecture would be a computable upper bound on the guessing probability:
\[ p_{\rm guess}(X|B) \leq e^{-Q(X|B)}. \]
We prove the conjecture for the case $n=2$ with pure states of $B$:
\begin{proposition}
If $\rho_{XB}= \sum_{x=1}^n p_x \proj{x} \otimes \rho^x_B$ has $n=2$ and $\rho^x_B$ are pure states then
\[ {Q(X|B)} \leq H_{\min} (X|B) .\]
\end{proposition}
\noindent {\bf Proof}\, (outline)  For the case of two pure states $\ket{\psi_1}$ and $\ket{\psi_2}$ with probabilities $p_1, p_2$ an explicit expression for the guessing probability has been given by Holevo and Helstrom \cite{hol,hel}. Writing $\cos \theta  =\bra{\psi_1} \psi_2\rangle$ we have
$p_{\rm guess} = \frac{1}{2}\left( 1+\sqrt{1-4p_1p_2\cos^2 \theta} \right)$, so
\[ H_{\min}(X|B) = -\ln \left(  \frac{1}{2}\left( 1+\sqrt{1-4p_1p_2\cos^2 \theta} \right) \right) .
\]
Also the eigenvalues of $\rho_B=p_1\proj{\psi_1}+p_2\proj{\psi_2}$ are $\lambda_{\pm}= \frac{1}{2}\left( 1\pm \sqrt{1-4p_1p_2\sin^2 \theta}\right)$ and
\[ Q(\rho_B)= \frac{1}{(\lambda_+-\lambda_-)} \left( \lambda_-^2\ln \lambda_-- \lambda_+^2 \ln \lambda_+ \right)  .\]
Note that $Q(\rho_{XB})$ is independent of $\theta$ (as the pure states are placed in orthogonal parts of the total $XB$ space by the $X$-register). Thus
\be\label{startap}
\frac{\partial\, Q(X|B)}{\partial \,\theta} = -\frac{\partial \,Q(\rho_B)}{\partial\, \theta}.
\ee
Viewing all quantities as functions of $\theta$ (for fixed $p_1,p_2$), 
it can be shown from the above expressions that
\be\label{algebra}
\frac{d}{d\theta} \left[H_{\min}(X|B) -Q(X|B)\right] \,\,\begin{array}{ll} <0 & \mbox{for $\theta \in (0,\pi/2) $}\\
>0 & \mbox{for $\theta \in (\pi/2, \pi)$}, \end{array}
\ee
as detailed in the Appendix.
Hence, $H_{\min}(X|B) -Q(X|B)$ has a  minimum at $\theta=\pi/2$. For this value of $\theta$, both terms are zero:  $H_{\min}(X|B)=0$ since $p_{\rm guess }(X|B)=1$, and
$Q(X|B)=0$ since  $Q(\rho_{XB})= Q(\rho_B)$, which follows from the fact that the eigenvalues of $\rho_{XB}$ and $\rho_B$ are identical for this value of $\theta$ ($\lambda_-=p_1$, $\lambda_+ = p_2$). Hence we deduce that $H_{\min}(X|B) -Q(X|B) \geq 0$ for all $\theta \in (0,\pi)$, as required.\, $\Box$

\subsection{Classical subentropy - an operational interpretation}
The function $F$ in eq.(\ref{qdef}) for subentropy may be applied to any probability distribution and it is interesting to ask if this expression has a significance in {\em classical} information theory.

The interpretation of subentropy $Q(\rho)$ given in \cite{jrw}, as a lower bound on the accessible information of any pure $\rho$-ensemble, appears not to have a direct classical information theoretic analogue. It is important here that the $\rho$-ensembles comprise {\em pure} states since if mixed states were allowed then for example, the $\rho$-ensemble comprising just $\rho$ itself with unit probability would have accessible information zero. The natural classical analogue of a pure state is a point mass probability distribution $\vc{\delta}_k = ( 0,\ldots ,0,1,0, \ldots ,0)$ (having 1 in the $k^{\rm th}$ slot) since, as for quantum states, arbitrary classical states (probability distributions) are then convex combinations of these, and they are themselves indecomposable. However, in contrast to the quantum case, any classical state (probability distribution)
has only a single {\em{unique}} decomposition as a mixture of pure states and the issue of {{minimal}} accessible information does not arise.  Nevertheless, classical subentropy can be given an operational interpretation as a lower bound on the capacity of a suitable class of classical channels, as described below.

Let $X$ denote a discrete random variable taking values in $\{1,2,\ldots, n\}$ with
probabilities $\vc{p}= (p_1, p_2, \ldots, p_n)$. Without loss of generality we assume that $p_1\ge p_2 \ldots \ge p_n$. We define the classical subentropy of $X$ as
\be\label{class_sub}
Q(X) := F(\vc{p})
\ee
where $F$ is the function in eq.(\ref{qdef}).

We first establish that the classical subentropy is equivalently given by an expression analogous to eq.\reff{rel1}, with the quantum relative entropy replaced by the Kullback-Leibler divergence, and the measurements $\cm$ replaced by a suitable class of stochastic maps described below. Let $\delta_i$ denote a probability vector of length $n$, with a $1$ only in the $i^{\rm th}$ slot, and let $P_\alpha$, $\alpha =1,2,\ldots, n$ denote  the $n$  cyclic permutation matrices of size $n\times n$. Then for every $i, \alpha \in \{1,2,\ldots, n\}$,
$$P_\alpha \delta_i = \delta_{(i+\alpha){{mod\, }}n}.$$
Consider the class of stochastic maps defined by
\be\label{cyclic}
\Lambda= \sum_{\alpha=1}^n t_\alpha P_\alpha,
\ee
where $\{t_\alpha\}_{\alpha =1}^n$ is a probability distribution. For any pair of probability distributions $\gamma=\{\gamma_i\}_{i=1}^n$ and $\mu=\{\mu_i\}_{i=1}^n$,
the Kullback-Leibler divergence is given by
$D(\gamma||\mu)= \sum_i \gamma_i \ln \left(\gamma_i/\mu_i\right)$. Writing
$p= \sum_{i=1}^n p_i \delta_i$, we have the following proposition.
\begin{proposition}\label{class_op}
The classical subentropy of a discrete random variable $X$ is given by
\be\label{lem}
Q(X) =  \sum_{i=1}^n p_i \langle D(\Lambda(\delta_i)|| \Lambda(p)\rangle_\Lambda,
\ee
where $\Lambda$ is the stochastic map in eq.(\ref{cyclic})
and $\langle \cdot \rangle_\Lambda$ denotes the uniform average over all such stochastic maps.
\end{proposition}

\noindent {\bf Remark}\,\, Recall that the quantum subentropy $Q(\rho )$ was expressed in terms of a uniform average over $\Delta_n$ resulting from the unitary Haar measure via Sykora's theorem. In Proposition \ref{class_op} we use a similar average but now coming from purely classical considerations.

\noindent {\bf Proof} \,
Setting $t_{ji} =t_{(j-i){{mod \,}}n}$
we obtain
\be
\left(\Lambda(\delta_i) \right)_j=t_{ji} \quad {\hbox{and}} \quad
\left(\Lambda(p) \right)_j=\sum_{i=1}^n t_{ji} p_i.
\ee
Hence
\begin{align}
\sum_{i=1}^n p_i \langle D(\Lambda(\delta_i)|| \Lambda(p))\rangle_\Lambda
&=  \langle \sum_{i=1}^n p_i \sum_{j=1}^n t_{ji} \ln t_{ji}
-  \sum_{i=1}^n p_i \sum_{j=1}^n t_{ji} \ln ( \sum_k t_{jk} p_k))\rangle_\Lambda\label{imp}\\
&=\langle I(X:\Lambda(X))\rangle_\Lambda \label{mut}
\end{align}
where $\Lambda(X)$ denotes a random variable taking values in $\{1,2,\ldots, n\}$  with probability distribution
$\{\sum_i t_{ji} p_i\}_{j=1}^n$, and the joint probability distribution
of $X$ and $\Lambda(X)$ is given by $\{t_{ji}p_i\}_{i,j=1}^n$. The uniform average over 
all stochastic maps of the form eq.\reff{cyclic} amounts to an average over the probability simplex $\Delta_n$
with respect to the normalized uniform measure $d{\mathbf t}$.

The second term on the right hand side of eq.\reff{imp} is the Shannon entropy of the random variable $\Lambda(X)$ averaged over the probability simplex.
In the first term, the averaging gives the same result for each summand. Hence, as in the proof of Proposition \ref{proprelent}, we see that the first
term equals $-C_n$ and the right hand side of eq.\reff{imp} reduces to
\be\label{fin}
n\int_{\Delta_n} \eta(\vc{p}.\vc{t})\,  d{\mathbf t} - C_n = F(\vc{p})\equiv Q(X),
\ee
completing the proof of the proposition.
Furthermore, eqs.\reff{mut} and \reff{fin} give
\be\label{ide2}
Q(X) = \langle I(X:\Lambda(X))\rangle_\Lambda
\ee
completing the proof. \,\, $\Box$

An operational interpretation of the classical subentropy $Q(X)$ can now be obtained from eq.\reff{ide2}.
Let $T$ be a multivariate random variable taking values $\vc{t}=(t_1,t_2,\ldots, t_n) \in \Delta_n$
with respect to the uniform distribution. Obviously, a particular choice of the stochastic map $\Lambda$ depends on the value $\vc{t}$ of $T$ and writing $\Lambda(X)\equiv \Lambda_T(X)$ we have $\langle I(X:\Lambda(X))\rangle_\Lambda$=$\langle I(X:\Lambda_T(X))\rangle_T$. Hence,
\begin{align}
Q(X)=\langle I(X:\Lambda(X))\rangle_\Lambda=\langle I(X:\Lambda_T(X))\rangle_T&= \int_{\Delta_n} d {\mathbf t}\,  I(X:\Lambda_t (X))\nonumber\\
&= \int_{\Delta_n} d{\mathbf t} \, I(X:\Lambda_T (X)|T=\vc{t})\nonumber\\
&=I(X:\Lambda_T(X)|T),\label{econd}
\end{align}
where the last quantity is the conditional mutual information.

Let $X$ denote the input to a classical channel $W$ which has two outputs: $T$ and $\Lambda_T(X)$ i.e. $W(X)=(T, \Lambda_T(X))$. By the chain rule
\be\label{cond_mutual}
I(X:\Lambda(X)|T) =  I(X:T,\Lambda(X)) - I(X:T) =  I(X:T,\Lambda(X)).
\ee
The last equality holds since $T$ is independent of $X$, so $I(X:T)=0$. From Lemma~\ref{class_op}
and eq.\reff{cond_mutual} we infer that
\be\label{expr}
Q(X) = I(X: W(X)).
\ee
Hence, for any discrete random variable $X$, the subentropy $Q(X)$ has the classical operational interpretation as the mutual information
between $X$ and the output $W(X)$ of the channel $W$ described above.
Moreover by Shannon's noisy channel coding theorem, the capacity $C(W)$ of the channel $W$ is given by:
\be\label{cap}
C(W) = \max_{\{p(x)\}} I(X:W(X)) =  \max_{\{p(x)\}} Q(X) \ge Q(X),
\ee
where the maximization is over all possible distributions for inputs to $W$.

\section*{Appendix: Proof of eq.\reff{algebra}}
Let us start with eq.\reff{startap} of Section~\ref{mincond}:
$$\frac{\partial\, Q(X|B)}{\partial \,\theta} = -\frac{\partial \,Q(\rho_B)}{\partial\, \theta}  = \frac{d Q(\rho_B)}{d \lambda_-} \, \frac{\partial \lambda_-}{\partial \theta}.$$
Writing $\lambda = \lambda_-$, we have
\begin{eqnarray*} \frac{d Q(\rho_B)}{d \lambda} &=& \frac{2 \lambda (1-\lambda)}{(1-2\lambda)^2}
\ln \frac{\lambda}{1-\lambda} + \frac{1}{1-2\lambda} \\ &=& -2 \int_0^1 (2x-1)
\left[ \ln (\lambda x + (1-\lambda)(1-x)) + 1 \right] dx \\
&=& \int_{-1}^1 u\,\ln [1+(1-2\lambda)u]\, du. \end{eqnarray*} Also
$$ \frac{\partial \lambda_-}{\partial \theta}
= \frac{p_1 p_2 \sin(2 \theta)}{\sqrt{1-4p_1 p_2 \sin^2(2\theta)}}
= \frac{p_1 p_2 \sin(2 \theta)}{1-2\lambda}. $$
It follows that \be -\frac{\partial}{\partial \theta} Q(X|B) = p_1
p_2 \sin(2 \theta) \frac{1}{s} \int_{-1}^1 u \ln(1+su) du, \ee
where $s = 1-2\lambda$. Now
$$ \frac{d}{ds} \int_{-1}^1 u \ln (1+su) du = \int_{-1}^1 \frac{u^2\, du}{1+su}
= 2 \int_0^1 \frac{u^2}{1-s^2 u^2} du $$ which is clearly increasing in $s$,
so that the integral is a convex function of $s \in [0,1]$.
Moreover
$$ \int_{-1}^1 u\,\ln(1+u)\,du = 1 \mbox{ and } \int_{-1}^1 u\,du = 0. $$
Therefore \be \int_{-1}^1 u\,\ln(1+su)\,du \leq s. \ee We conclude
that \be -\left(\frac{\partial}{\partial \theta} H_{\min}(X|B)
-\frac{\partial}{\partial \theta} Q(X|B) \right) > 0 \ee for
$\theta \in (0,\pi/2)$ and $<0$ for $\theta \in (\pi/2,\pi)$.
\subsection*{Acknowledgments}
We thank Graeme Mitchison, Sergii Strelchuk and Andreas Winter for helpful suggestions and discussions.

\end{document}